\documentstyle[12pt,psfig]{article} 
\newcommand{\ber}{\begin{eqnarray}}
\newcommand{\eer}{\end{eqnarray}}
\newcommand{\bea}{\begin{equation}}
\newcommand{\eea}{\end{equation}}

\def\gtap{\raisebox{-.4ex}{\rlap{$\sim$}} \raisebox{.4ex}{$>$}}

\begin{document}

\begin{flushright}
November 2002\\
hep-ph/0303229
\end{flushright}
\vskip 50pt
\begin{center}
{\Large \bf A note on the neutrino mass implications of the K2K experiment}\\
\vspace{1cm}
\renewcommand{\thefootnote}{\fnsymbol{footnote}}
{\large \sf Suvadeep Bose$^{1,\!}$\footnote{E-mail address:
sb416@cam.ac.uk}},
{\large \sf Amitava Raychaudhuri$^{2,\!}$\footnote{E-mail
address: amitava@cubmb.ernet.in}}
\vskip 10 pt
$^{1}${\small \it St Edmund's College, University of Cambridge,
Cambridge CB3 0BN, U.K.}\\
$^{2}${\small \it Department of Physics, University of Calcutta,\\
92 Acharya Prafulla Chandra Road, Kolkata 700009, India}
\normalsize
\end{center}

\begin{abstract}
{\small 
The K2K experiment has presented the first results on the
observation of $\nu_\mu$. They show a depletion compared to
the expectations and are consistent with neutrino oscillations with a
mass-splitting in the range favoured by the Super-Kamiokande
atmospheric neutrino measurements. Here we examine the extent by which
the range of $\Delta m^2$ obtained from the K2K measurements can vary
due to the uncertainties in the flux, cross-section, and detector
efficiency.
} 
\vskip 5 pt 
\end{abstract}

\vskip 20pt

\centerline{\texttt{PACS Nos. 14.60.Pq, 14.60.Lm }}

\vskip 10pt

\centerline{Short title: Neutrino mass implications of the K2K experiment}

\vskip 10 pt

\renewcommand{\thesection}{\Roman{section}}
\setcounter{footnote}{0}
\renewcommand{\thefootnote}{\arabic{footnote}}

\section{Introduction}

A non-zero neutrino mass has far-reaching implications in particle
physics, astrophysics, and cosmology. Any evidence in its support
must be probed from as many different angles as possible.
Atmospheric neutrinos provide a strong case for non-degenerate
neutrino masses and mixing.  Super-Kamiokande (SK) \cite{sk},
MACRO \cite{macro}, and Soudan 2 \cite{soudan}  studying neutrinos
produced in interactions of cosmic rays with atmospheric nuclei
show a significant suppresion of the ratio of $\nu_{\mu}$ to
$\nu_{e}$ with respect to the expectation from standard
hadronic shower models. These observations can be explained by the
phenomenon of neutrino oscillations in which $\nu_{\mu}$ changes
to some other neutrino, {\em e.g.}, $\nu_\tau$. Such oscillations
can occur if neutrinos have a non-degenerate mass spectrum and if
the neutrino mass eigenstates are not the same as the flavour
eigenstates. The atmospheric neutrino data favour a mass
splitting $\Delta m^2 = |m_1^2 - m_2^2| \simeq 10^{-3}$ eV$^2$
and maximal mixing, $\theta = \pi/4$.  Neutrino oscillations have
also been advanced as a natural solution to the long-standing
solar neutrino problem, especially when the data from the SK and
SNO experiments are taken together.

K2K \cite{k2k} is the first accelerator based long-baseline
oscillation experiment. It is designed precisely to probe the
mass and mixing parameters favoured by the atmospheric
neutrino data. It  uses neutrinos produced by the 12 GeV proton
beam of the KEK accelerator laboratory which are subsequently
detected by the SK detector at a distance of 250 km.  The average
neutrino energy is $\langle E \rangle =1.3$ GeV and the intense beam is
$98.0\%$ pure $\nu_{\mu}$.  K2K also uses a near detector at KEK, 300m
downstream from the pion production target, to monitor and study the
neutrino beam. The experiment started taking data from June, 1999.
The primary oscillation search mode is $\nu_{\mu}
\rightarrow \nu_{x}$ oscillation (the $\nu_{\mu}$ disappearance
mode).

The first results from K2K \cite{k2k2} have been released. With
3.85 $\times 10^{19}$ protons on target (p.o.t.), in the
22.5 kton fiducial volume of the SK detector they observe 44
fully contained events in place of the (no oscillation)
expectation of $63.9^{+6.1}_{-6.6}$ events. This corresponds to a
depletion ratio 
\begin{equation}
d \equiv \frac{\rm No.~of~observed~events}{\rm No.~of~expected
~events} = 0.689^{+0.079}_{-0.060} \cdot
\label{eq:depl}
\end{equation}
This depletion is consistent with neutrino oscillations with a
mass splitting in the atmospheric range and maximal
mixing\footnote{Based on 2.6$\times 10^{19}$ p.o.t. \cite{k2ko}
the depletion ratio was $d = 0.670^{+0.086}_{-0.070}\cdot$}. In
this note, basing ourselves on the available information, we wish
to examine to what extent this conclusion could be uncertain. Our
finding is that the result is robust.

When oscillations are operative, the survival probability of a
neutrino with energy $E$ produced a distance $L$  from a
detector is expressed (in the two-flavour approximation) as:

\bea
P_{\mu\mu} = P(\nu_{\mu} \rightarrow \nu_\mu) = 1 - \sin^{2}2 \theta
\sin^{2}\frac{1.27 \Delta m^{2}({\rm eV}^{2})L({\rm km})}{E({\rm
GeV})},
\label{eq:prob}
\eea

For the purpose of the present work, the nature of the partner
neutrino state to which the $\nu_\mu$ oscillates is immaterial so
long as it does not mimic a $\nu_\mu$ at the SK detector.

\section{The K2K results}

In the presence of neutrino oscillations, the event rate in the
K2K experiment, $r$, can be expressed as

\bea
r = \int dES(E)P_{\mu\mu}(E),
\label{eq:r}
\eea
where $P_{\mu\mu}$ is given by (\ref{eq:prob}).

The oscillation parameters $\theta$ and $\Delta m^{2}$ will be 
determined from the K2K data.  The experiment is sensitive to
$(\Delta m^{2},\sin^22\theta)$ around $(10^{-3}$eV$^{2},1)$ favoured
by the SuperKamiokande experiment.

$S(E)$ is the `energy spectrum' of the neutrinos from KEK  
as detected at the Kamioka site. This spectrum is
given by the convolution of the standard (no oscillation)
neutrino energy spectrum $\phi$ at the SK (far) detector,
the interaction cross-section $\sigma$, and the
detection efficiency $\epsilon$ (all in differential form),
integrated over the final state parameters, {\em i.e.}, symbolically,
$S = \int \phi \sigma \epsilon$.

In the published literature there is information on $\phi$ but
not yet on the product $\sigma . \epsilon$ relevant for the far
detector specification. As direct and accurate reconstruction of
the spectrum is not possible, some indirect, approximate
reconstruction was suggested by the K2K collaboration.

We have two sets of information in hand.

(i) The neutrino spectrum is considered to be practically zero above 5
GeV \cite{k2k}.

(ii) According to the K2K Monte Carlo simulation, for an exposure of
$3.85\times10^{19}$ p.o.t., assuming maximal mixing, we have the set of
predictions for the event rate ($r^{MC}$) for different $\Delta m^{2}$
shown in Table 1:

\vskip 5 pt
\begin{center}
\begin{tabular}{|c|c|c|c|c|}
\hline
$\Delta m^{2}$ in eV$^{2}$ & $1\times10^{-4}$ & $3\times10^{-3}$ &
$5\times10^{-3}$ & $7\times10^{-3}$\\
\hline
$r^{MC}$ & $63.9$ & $41.5$ & $27.4$ & $23.1$\\
\hline
\end{tabular}
\end{center}
\begin{description}
\item{\small \sf Table 1:} {\small \sf  The values of $r^{MC}$
for different $\Delta m^2$ determined from the K2K Monte Carlo
calculations \cite{k2k2} for maximal mixing ($\sin^22 \theta = 1$).}  
\end{description}

\section{Impact of different $S(E)$}
In the absence of direct information, a form for the spectrum
was suggested \cite{flm} to be $S(E) \propto
x^{\alpha}(1-x)^{\beta}$ with $x = E/($5 GeV). It fits the expected
behaviour of $S(E)$ at $x$ = 0 and 1 and has a particularly
simple form. The overall normalization and the parameters
$\alpha$ and $\beta$ were obtained by using a best fit to the
data in Table 1.  It was shown using this functional form of
$S(E)$ that the preliminary data from K2K are consistent with the
$\Delta m^2$ favoured by the SK atmospheric neutrino experiment.

Here, we ask the question: How sensitively does this conclusion
depend on the form of $S(E)$? To this end, we consider several
different functional forms for the spectrum $S(E)$. For each, the
best-fit values of the parameters are determined by minimizing
\bea
\chi^2 = \sum_{i=2}^4 \frac{(r_i - r_i^{MC})^2}{\sigma_i^2},
\eea
where $r_i$ is calculated using the chosen form of $S(E)$ and
$r_i^{MC}$ is taken from the K2K Monte Carlo results of Table 1
and, as in \cite{flm},
\bea
\sigma_i^2 = 44 + 0.01 r_i^2.
\label{eq:sigma}
\eea
$r^{MC}$ corresponding to $i=1$, {\em i.e.}, $\Delta m^2 = 1.0 \times
10^{-4}$ eV$^2$, is used to determine the overall normalization
constant $N$. In this way, in all cases the no oscillation limit is
correctly reproduced. The functional forms that we have examined and
the best-fit values of the parameters are listed in Table 2.  These
functions are plotted in Fig.~1.

\begin{center}
\begin{tabular}{|c|c|c|c|c|c|}
\hline
Case& functional & Normalization & \multicolumn{3}{|c|}{best fit
values} \\ \cline{4-6}
&form&$N$ & $\alpha$ & $\beta$ & $\chi^2$\\
\hline
$a$&$N x^{\alpha}(1-x)^{\beta}$ & 1370.7 & 1.14 & 2.78 & 0.044\\
$b$&$N x^{\alpha}(1-x^{2})^{\beta}$ & 518.3 & 0.85 & 3.81 & 0.063\\
$c$&$N x^{\alpha}(1-x^{3})^{\beta}$ & 307.7 & 0.60 & 5.50 & 0.068\\
$d$&$N x^{\alpha}(1-x^{4})^{\beta}$ & 215.7 & 0.42 & 7.48 & 0.061\\
\hline
\end{tabular}
\end{center}
\begin{description}
\item{\small \sf Table 2:} {\small \sf  The different
parametrizations of $S(E)$, the best-fit values of the parameters,
and the values of $\chi^2$.}
\end{description}

All the above functions have a common nature in that they vanish at
$x=$ 0 and 1 and rise to a maximum value at some intermediate energy.
The actual neutrino spectrum as a function of the energy is expected to
have  rising and falling regions at the two ends and an intermediate
region where its variation is rather slow. To mimic this feature, we
have also considered two further cases where we have chosen the
functional forms of Cases $a$ and $b$ of Table 2 but `chopped' the
function at the values of $E$ where it achieves half the maximum value.
Between these points, $E_1$ and $E_2$, the function is assumed to be a
constant. The points $x_i = E_i$/(5 GeV), ($i$ =1,2) and the best-fit
values of the parameters in these cases are presented in Table 3 and
the functions exhibited in Fig.~1.

\begin{center}
\begin{tabular}{|c|c|c|c|c|c|c|c|}
\hline
Case&functional & Normalization &
\multicolumn{2}{|c|}{Half-Maxima} & \multicolumn{3}{|c|}{best fit
values} \\ \cline{4-8}
& form & $N$ & $x_1$ & $x_2$ &$\alpha$ & $\beta$ & $\chi^2$\\
\hline
$e$&$N x^{\alpha}(1-x)^{\beta}$ & 13136.3&0.128 & 0.529 &
2.00&4.61 & 0.065 \\ \hline
$f$&$N x^{\alpha}(1-x^{2})^{\beta}$ & 3392.6 & 0.156 & 0.544 &
1.72 & 6.58 & 0.047 \\
\hline
\end{tabular}
\end{center}
\begin{description}
\item{\small \sf Table 3:} {\small \sf  
Parametrizations of $S(E)$ with a constant region between $x_1$
and $x_2$ (see text). The best-fit values of the parameters
and the values of $\chi^2$ are shown.}
\end{description}

We expect that the six rather different functional forms for
$S(E)$ that we have chosen adequately capture the possible
uncertainties in the neutrino spectrum, cross-section, and
detection efficiency.  We make two remarks. Firstly, the fits are
all of very low $\chi^2$. This is due to the comparatively large
$\sigma_i^2$ from (\ref{eq:sigma}) associated with each datum
point. Secondly, we find that over a rather broad region in the
($\alpha, \beta$) parameter space around the best-fit points,
$\chi^2$ varies relatively little. In particular, the best-fit
point we find in Case $a$ is somewhat different from the one in
\cite{flm} even though the functional forms are the same. For the
latter we find a comparable but higher $\chi^2$.

After having obtained the six different functional forms of
$S(E)$ which best fit the Monte Carlo results of Table 1 we use
them in (\ref{eq:r}) to obtain the predicted value for the number
of events for the K2K experiment with $3.85\times10^{19}$ p.o.t.
These results are shown in Fig. 2 as a function of $\Delta m^2$
for the maximal mixing case ($\sin^22\theta$ = 1). The two
horizontal lines are the 1$\sigma$ range of the experimental
observation. The intercepts of the curves with this experimental
range determine the allowed values of $\Delta m^2$ for the
various cases. It is seen from Fig. 2 that the differences
between the alternative cases $(a$ - $f)$ are comparatively marginal.
There is a small spread in the lowest admissible values of
$\Delta m^2$ in the different cases. But as the data improve even
this will be removed.  Since we have chosen $S(E)$ with a wide
range of forms, it would not be unfair to conclude that the range
of $\Delta m^2$ which is favoured by the K2K data can be
predicted in a robust fashion.

Motivated by the best-fit to the atmospheric neutrino data, in
this work we have used maximal mixing, {\em i.e.},
$\sin^22\theta = 1.0$, in (\ref{eq:prob}). The SK data restrict
the mixing angle rather tightly and at 90\% C.L. one can allow
$\sin^22\theta \gtap 0.9$. How much would our conclusions be
affected if the lower limit of $\sin^22\theta$ is used in the analysis?
We find, for example, for an $S(E)$ of the form of Case $a$ (see Table
2) the best fit point moves to ($\alpha, \beta$) = (2.39,
6.10)\footnote{This fit has $\chi^2 = 9.74 \times 10^{-3}$.}. The event
rate as a function of $\Delta m^2$ for this case is also shown in Fig.
2. Notice that a small region around $2 \times 10^{-2}$ eV$^2$ is
allowed for this case.  A large improvement in the K2K systematic and
statistical uncertainties will be able to exclude this option.

\section{Conclusion}

The first results from the K2K long baseline experiment show a
depletion in the number of events compared to the expectation.
This is consistent with oscillations of the $\nu_\mu$ to a
neutrino of a different flavour with mass and mixing in the range
favoured by the atmospheric neutrino results. In this
note, we have shown that this conclusion is robust and is not
altered even if the uncertainty in the initial neutrino spectrum,
interaction cross-section, and detection efficiency are widely
varied.  

{\em Note added:} After the submission of the paper for
publication, K2K announced results \cite{k2kn} based on $4.8
\times 10^{19}$ p.o.t. They observe 56 events with a
no-oscillation expectation of 80.1$^{+6.2}_{-5.4}$. This
corresponds to the depletion ratio (see Eq.~(\ref{eq:depl})) $d =
0.699^{+0.051}_{-0.050}$, consistent with the earlier result. We
find using our analysis that for the different parametrizations
of $S(E)$ this corresponds to allowed ranges for $\Delta m^{2}$
shown in Table 4. It is gratifying that the best fit point found
by the K2K group for maximal mixing ($\Delta m^{2} = 2.8 \times
10^{-3}$ eV$^2$ \cite{k2kn}) lies within the allowed range for all
parametrizations considered.

\begin{center}
\begin{tabular}{|c|c|c|}
\hline
Case& \multicolumn{2}{|c|}{$\Delta m^{2}$ Limits (10$^{-3}$ eV$^2$)}
\\ \cline{2-3}
&Lower& Upper \\
\hline
$a$& 1.57 & 3.61 \\ \hline
$b$&1.60 & 3.72 \\ \hline
$c$&1.52 & 3.40 \\ 
   &22.48&25.06 \\ \hline
$d$&1.47 & 3.73 \\ 
   &21.77& 26.38\\ \hline
\end{tabular}
\end{center}
\begin{description}
\item{\small \sf Table 4:} {\small \sf  The ranges of $\Delta
m^{2}$ allowed at 1$\sigma$ for maximal mixing by the latest K2K
results\cite{k2kn} for the alternative parametrizations of $S(E)$ (see
Table 2). Notice that in cases (c) and (d) there are two allowed
ranges.}
\end{description}

\section*{Acknowledgements}
SB has been partially supported by St. Edmund's College,
Cambridge, U.K., while the work of AR has been supported in part
by C.S.I.R., India and D.S.T., India.

\newpage
\vskip 30pt
\begin{figure}[thb]
\vskip -1.00in
\psfig{figure=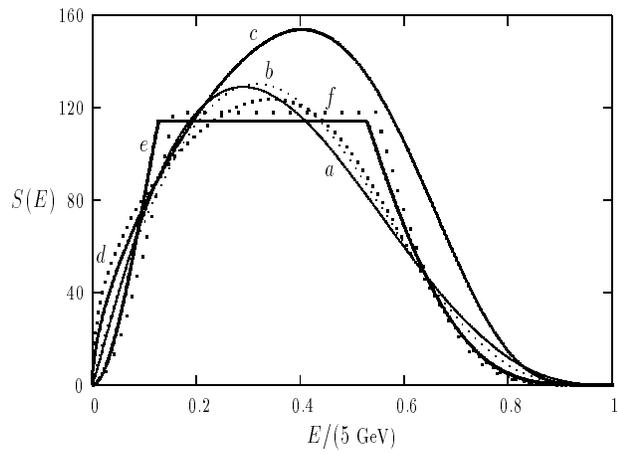,width=14.0cm,height=22.0cm} 
\vskip -5.10in
\caption{\sf \small   
The  spectrum $S(E)$ as a function of  $E$. The cases to which
the different curves correspond are indicated   (see Tables 2 and
3). }

\end{figure}

\newpage
\vskip 30pt
\begin{figure}[thb]
\vskip -1.00in
\psfig{figure=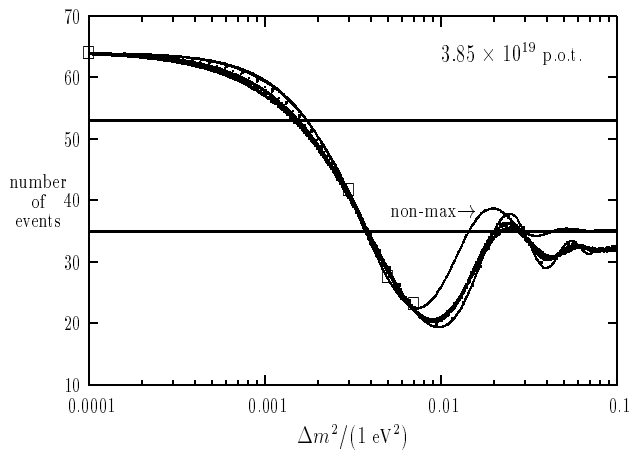,width=14.0cm,height=22.0cm} 
\vskip -5.10in
\caption{\sf \small   
The number of events in the K2K experiment for an exposure of
$3.85\times 10^{19}$ p.o.t. as a function of $\Delta m^{2}$ (at
$\sin^22\theta =1)$.  The region between the two horizontal lines is
the current K2K measured range within 1 standard deviation.  The boxes
are the values from the K2K MC simulation (Table 1). The six curves
obtained from the  different forms for $S(E)$ considered in this work
are shown (same conventions as in Fig. 1). The curve marked `non-max'
corresponds to the case of non-maximal mixing ($\sin^22\theta = 0.9$)
discussed in the text.}
\end{figure}

\end{document}